\documentclass[11pt,a4paper]{article}

\usepackage{geometry}
\usepackage[applemac]{ inputenc}

\usepackage{mathrsfs}
\usepackage{amsmath}
\usepackage{amssymb}
\usepackage{multicol}
\usepackage{float}
\usepackage{makeidx}
\usepackage{layout}
\usepackage{array}
\usepackage{hyperref}
\usepackage{latexsym}
\usepackage{indent first}

\usepackage{subcaption,graphicx}
\usepackage{appendix}

\usepackage{color}

\usepackage[sectionbib,square]{natbib}

\usepackage[normalem]{ulem}

\title{Gravity-capillary wave interactions generated by moving disturbances: Euler equations framework}
\author{M. V. Flamarion$^{1}$ and R. Ribeiro-Jr$^{2}$}

\date{}

\begin{document}

\maketitle

{\footnotesize
	\begin{center}
	$^{1}$ UFRPE/Rural Federal University of Pernambuco, UACSA/Unidade Acad{\^e}mica do Cabo de Santo Agostinho, BR 101 Sul, 5225, 54503-900, Ponte dos Carvalhos, Cabo de Santo Agostinho, Pernambuco, Brazil.
	
		$^2$ UFPR/Federal University of Paran\'a,  Departamento de Matem\'atica, Centro Polit\'ecnico, Jardim das Am\'ericas, Caixa Postal 19081, Curitiba, PR, 81531-980, Brazil. 
	
	\end{center}
	
	}
\begin{abstract}
The aim of this work is to investigate the interaction of gravity-capillary waves resonantly excited by two moving disturbances along the free surface.  The problem is formulated using the full Euler equations and numerical computations are performed in a simplified domain through the use of a conformal mapping which   flattens the free surface. We focus on nearly-critical flows with intermediate capillary effects and characterise their main features. In the supercritical regime  the wave interaction can be strongly nonlinear leading to the onset of wave breaking. In the subcritical regime depression solitary waves are generated remaining  trapped between the disturbances  bouncing back and forth. In addition, we notice a dependence of the number of trapped waves on the distance of the disturbances. Furthermore, differently from when only on disturbance is considered, we find that the critical regime captures features from the subcritical and supercritical regime simultaneously.
  \\

{\bf Key words:} Gravity-capillary waves, Solitary waves, Trapped waves, Euler equations.
 
\end{abstract}

\section{Introduction}

Waves excited by a moving disturbance have been studied using different mathematical models. The main framework used is the full Euler equations (\cite{Grimshaw13, Hirata, Marcelo-Paul-Andre}). However, due to intrinsic difficulties present in the Euler equations such as nonlinearity and free boundary conditions, reduced models based on asymptotic theory have been applied as an alternative to describe this phenomenon. The most notorious are the forced Korteweg-de Vries equation (fKdV) (\cite{Wu2, Akylas, Grimshaw86, Wu1, Paul, Milewski3, M-Selecciones}) and the fifth-order fKdV equation (\cite{Milewski3, Zhu, Capillary}).

The flow is governed by three parameters: the magnitude  of the applied pressure, the Froude number ($F$) and the Bond number ($B$) defined as
$$F = \frac{U_{0}}{\sqrt{gh_{0}}},\;\ B = \frac{\sigma}{\rho g h_{0}^2}.$$
Here, $U_0$ is the speed of the applied pressure forcing, $g$ is the acceleration of gravity, $h_{0}$ is the undisturbed depth of a  channel, $\sigma$ is the coefficient of the surface tension and $\rho$ is  the constant density of the fluid. The Froude number and the Bond number are called critical when $F= 1$  and $B=1/3$. In the weakly nonlinear regime the fKdV equation is applied to study nearly-critical flows ($F\approx 1$) with external forcings of small amplitudes (\cite{Wu2, Akylas, Grimshaw86, Wu1, Paul}), however it fails when $B=1/3$ since the higher order dispersive term vanishes and the fKdV approaches asymptotically to a Burguers equation (\cite{Falcon}). When the flow is nearly-critical and the capillary effects are intermediate ($B\approx1/3$) higher order fKdV models arise. Under these conditions, \cite{Milewski3} derived a fifth-order fKdV for an obstacle of small amplitude and showed that  this equation has unsteady solitary wave solutions with small oscillating tails. More recently, \cite{Hirata} used the body-fitted curvilinear coordinates to solve Euler's equations numerically in the presence of an obstacle with a uniform flow and compared the results with the fKdV and the  fifth-order-fKdV for the critical flow and intermediate capillary effects ($B\approx 1/3$). They showed that the fifth-order-fKdV qualitatively captures the main features of the flow, however it overestimates the wavelengths.

Considering two well separated localised obstacles in the absence of surface tension, \cite{Grimshaw16} used the fKdV model to investigate the interaction of the  waves generated in the nearly-critical regime. According to their results the flow is classified into three steps. In the first step, which occurs at early times, the flow is featured by the formation of an undular bore above each obstacle independently. In the second step the waves generated  interact between the obstacles, and in the third one happens the controlling of the dynamic by the larger obstacle. Later, these authors revisited the interaction of  waves generated over two obstacles (bumps and holes) in the nearly-critical regime and detailed the wave interactions (\cite{Grimshaw19}).  More recently, \cite{Capillary} investigated gravity-capillary flows over obstacles and showed that different from gravity waves the flow is not necessarily governed by the obstacle with larger amplitude. Moreover, they found evidences that in the subcritical regime ($F<1$) waves may be trapped between the obstacles. Although there are results on flows with multiple external forcings using reduced models, as far as we know there are no articles on this topic concerning Euler equations.

In this paper we use the full Euler equations to  investigate numerically  the interaction of gravity-capillary waves excited  by two disturbances moving along the free surface with nearly-critical speed ($F\approx 1$) in a finite depth channel under  intermediate capillary effects ($B=1/3$).  The problem is formulated in the disturbances moving frame and computations are performed in a simplified domain through the use of the conformal mapping technique. We find that in the supercritical regime ($F>1$) the wave interaction is strongly nonlinear and numerical evidences show that the wave may break. In the subcritical regime ($F<1$) ,  we observe the generation of depression solitary waves which bounce back and forth between the disturbances remaining trapped for large times. We notice that the larger is the distance of the disturbances the larger is the number of trapped waves. Besides, in the critical regime ($F=1$), we observe a formation of an undular bore where the pressure is applied and the arising   of depression solitary waves for large times which does not occur when only one disturbance is considered. It is worthy to mention that all these features have not been reported in previous works.

This article is organized as follows. In section 2 we present the mathematical formulation of the non-dimentional Euler equations.  The conformal mapping formulation of the problem and numerical methods are presented in section 3, the results in section 4 and the conclusion in section 5.

\section{Mathematical Formulation}
We consider an inviscid fluid with constant density ($\rho$) in a  two-dimensional channel with finite depth ($h_0$) under the force of gravity ($g$), with surface tension ($\sigma$), and in the presence of a left-going pressure distribution ($P$) which travels with constant speed ($U_0$) along the free surface.  The flow is assumed to be incompressible and irrotational. Moreover, we denote the velocity potential in the bulk fluid by $\tilde\phi (x, y, t)$ and the free surface by $\tilde{\zeta}(x, t)$. Using the typical wavelength $h_0$ as the horizontal and vertical length, $(gh_0)^{1/2}$ as the velocity potential scale, $(h_0/g)^{1/2}$ as the time scale, and $\rho g h_0$ as the pressure scale, we obtain the dimensionless Euler equations 
\begin{align} \label{eu2}
\begin{split}
& \Delta{\tilde{\phi}}= 0 \;\  \mbox{for} \;\ -1 < y <{\tilde{\zeta}}(x,t), \\
& {\tilde{\phi}}_{y} =0\;\ \mbox{at} \;\ y = -1, \\
& {\tilde{\zeta}}_{t}+{\tilde{\phi}}_{x}{\tilde{\zeta}}_{x}-{\tilde{\phi}}_{y}=0
\;\ \mbox{at} \;\ y = {\tilde{\zeta}}(x,t), \\
& {\tilde{\phi}}_{t}+\frac{1}{2}({\tilde{\phi}}_{x}^2+{\tilde{\phi}}_{y}^{2})+{\tilde{\zeta}} -B\frac{\tilde{\zeta}_{xx}}{(1+\tilde{\zeta}_x^2)^{3/2}}= - P(x+Ft) \;\ \mbox{at} \;\ y = {\tilde{\zeta}}(x,t),
\end{split}
\end{align}
where $F=U_{0}/(gh_0)^{1/2}$ is the Froude number and $B=\sigma/\rho g h_{0}^2$ is the Bond number. Now, we rewrite the equations (\ref{eu2}) in the moving frame  $x\rightarrow x+Ft$. To this end, we define
\begin{equation}\label{moving}
\tilde{\zeta}(x-Ft,t)=\bar{\zeta}(x,t), \;\ \tilde{\phi}(x-Ft,y,t)=\bar{\phi}(x,y,t).
\end{equation}
Substituting (\ref{moving}) in (\ref{eu2}) yields the system 
\begin{align} \label{eu3}
\begin{split}
& \Delta{\bar{\phi}}= 0 \;\  \mbox{for} \;\ -1 < y <{\bar{\zeta}}(x,t), \\
& {\bar{\phi}}_{y} =0\;\ \mbox{at} \;\ y = -1, \\
& {\bar{\zeta}}_{t}+F{\bar{\zeta}}_{x}+{\bar{\phi}}_{x}{\bar{\zeta}}_{x}-{\bar{\phi}}_{y}=0
\;\ \mbox{at} \;\ y = {\bar{\zeta}}(x,t), \\
& {\bar{\phi}}_{t}+F{\bar{\phi}}_{x}+\frac{1}{2}({\bar{\phi}}_{x}^2+{\bar{\phi}}_{y}^{2}) +{\bar{\zeta}}-B\frac{\bar{\zeta}_{xx}}{(1+\bar{\zeta}_x^2)^{3/2}}= - P(x) \;\ \mbox{at} \;\ y = {\bar{\zeta}}(x,t).
\end{split}
\end{align}
In the next section we write the equations (\ref{eu3}) in a simplified domain using a conformal mapping and present the numerical method to solve them.

\section{ Conformal mapping and numerical methods}
We solve the system (\ref{eu3}) using the method introduced by \cite{Dyachenko}. Here, we summarise only the main steps. First, we construct a time-dependent conformal mapping 
\begin{equation*}
z(\xi,\eta,t) = x(\xi,\eta,t)+iy(\xi,\eta,t),
\end{equation*}
which flattens the free surface and maps a strip of width $D(t)$ onto the fluid domain and satisfies the boundary conditions 
\begin{equation*}
y(\xi,0,t)=\overline{\zeta}(x(\xi,0,t),t) \;\ \mbox{and} \;\ y(\xi,-D(t),t)=-1.
\end{equation*}
\begin{figure}[!ht]
	\centerline{\includegraphics[scale=0.4]{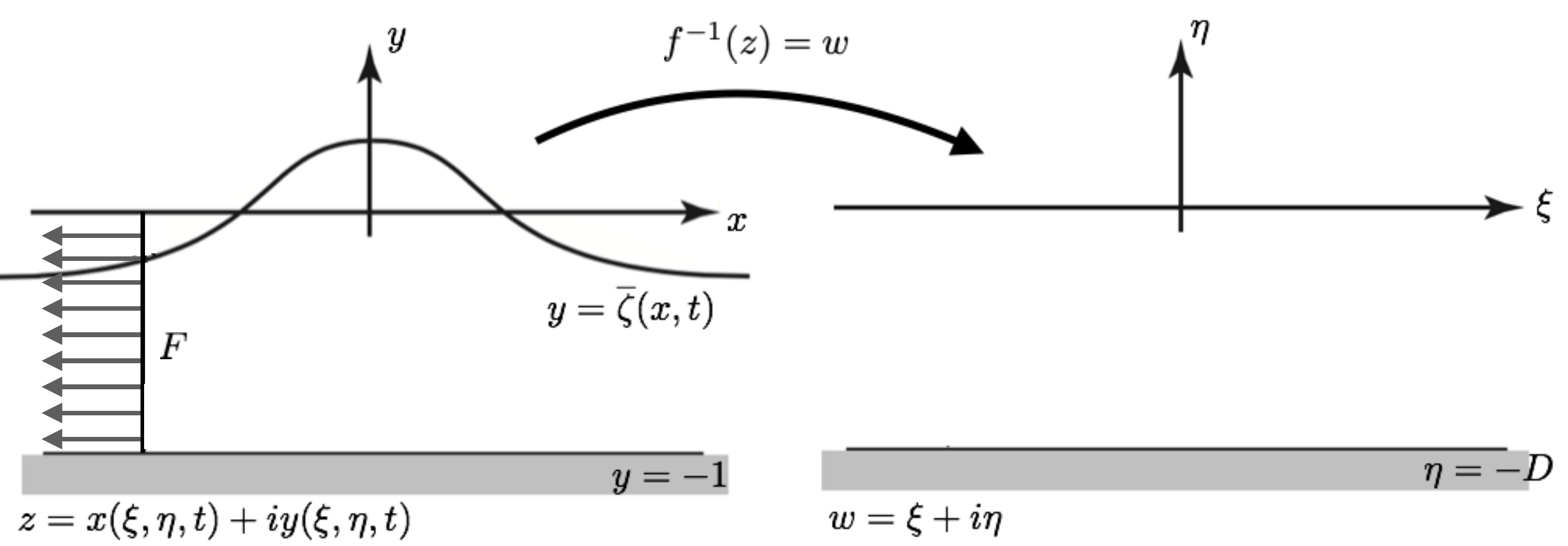}}
	\caption{Description of the conformal mapping. The free surface is flattened out in the canonical domain. }
	\label{e001e0005}
\end{figure}


Let $\phi(\xi,\eta,t)=\bar{\phi}(x(\xi,\eta,t),y(\xi,\eta,t),t)$ and $\psi(\xi,\eta,t)=\bar{\psi}(x(\xi,\eta,t),y(\xi,\eta,t),t)$ be its harmonic conjugate, and  denote by
$\mathbf{\Phi}(\xi,t)$ and $\mathbf{\Psi}(\xi,t)$ their traces along $\eta=0$ respectively. Considering $\mathbf{X}(\xi,t)$ and $\mathbf{Y}(\xi,t)$ as the horizontal and vertical coordinates at $\eta=0$, Kinematic and Bernoulli conditions $(\ref{eu3})_{3,4}$ are read as
\begin{align}\label{eulerconforme}
\begin{split}
& \mathbf{Y}_{t} =\mathbf{Y}_{\xi}\mathcal{C}\bigg[\frac{\mathbf{\Theta}_{\xi}}{J}\bigg] 
-\mathbf{X}_{\xi}\frac{\mathbf{\Theta}_{\xi}}{J}, \\
& \mathbf{\Phi}_{t} = - \mathbf{Y} - \frac{1}{2J}
(\mathbf{\Phi}_{\xi}^{2}-\mathbf{\Psi}_{\xi}^{2}) +\mathbf{\Phi}_{\xi}\mathcal{C}\bigg[\frac{\mathbf{\Theta}_{\xi}}{J}\bigg] 
- \frac{1}{J}F\mathbf{X}_{\xi}\mathbf{\Phi}_{\xi}+B\frac{\mathbf{X}_{\xi}\mathbf{Y}_{\xi\xi}-
\mathbf{Y}_{\xi}\mathbf{X}_{\xi\xi}}{J^{3/2}} - P(\mathbf{X}),
\end{split}
\end{align}
where $\mathbf{\Theta}_{\xi}(\xi,t)=\mathbf{\Psi}_{\xi}+F\mathbf{Y}_{\xi}$, $J=\mathbf{X}_{\xi}^2+\mathbf{Y}_{\xi}^2$ is the Jacobian of the conformal mapping evaluated at $\eta=0$, 
\begin{align*}
\begin{split}
& \mathbf{X}_{\xi} = 1-\mathcal{C}\big[\mathbf{Y}_{\xi}\big], \\
& \mathbf{\Phi}_{\xi} = -\mathcal{C}\big[\mathbf{\Psi}_{\xi}\big], \\
\end{split}
\end{align*}
and $\mathcal{C}$ is the operator  $$\mathcal{C}=\mathcal{F}^{-1}_{k\ne 0}i\coth(k_jD)\mathcal{F}_{k\ne 0},$$ 
where $\mathcal{F}$ denote the Fourier modes  $$\mathcal{F}_{k_j}[g(\xi)]=\hat{g}(k_j)=\frac{1}{2L}\int_{-L}^{L}g(\xi)e^{-ik_j\xi}\,d\xi,$$
$$\mathcal{F}^{-1}_{k_j}[\hat{g}(k_j)](\xi)=g(\xi)=\sum_{j=-\infty}^{\infty}\hat{g}(k_j)e^{ik_j\xi},$$
where $k_j=(\pi/L)j$, $j\in\mathbb{Z}$.
Computationally, it is interesting to impose the canonical domain and physical domain to have same length. To this end, we define $$D = 1+ \frac{1}{2L}\int_{-L}^{L}\mathbf{Y}(\xi,t) d\xi.$$
More details of this numerical method can be found in  \cite{Marcelo-Paul-Andre}. 

The dynamic of the waves generated is found integrating in time the family of ordinary differential equations (\ref{eulerconforme}) with the forth-order Runge-Kutta method and the derivatives in $\xi$ are  performed using the Fast Fourier Transform (FFT) (\cite{Trefethen}). In our simulations we take the computational domain $[-L,L]$, with a uniform grid with $N$ points where $L$ is taken large enough to avoid effects of spatial periodicity.
\section{Numerical results}
The disturbances over the free surface are modelled by two well separated  gaussians 
\begin{equation}\label{obstacles}
P(x) = \epsilon_{1}\exp\Big({-(x-x_a)^2/w}\Big)+\epsilon_{2} \exp\Big({-(x-x_b)^2/w}\Big),
\end{equation}
where $\epsilon_1$ and $\epsilon_2$ are the amplitudes of each disturbance,  $w$ is their width, and $x_a$ and $x_b$ are their locations. In order to see the contribution of each disturbance to the dynamic independently, we take the disturbances to be far apart, thus wave interactions only occur for large times. For this purpose, we fix  $x_{a}=-150$, $x_b=150$ and $w=100$.  In this paper we do not attempt an exhaustive study of the wave interactions. Instead we present interesting regimes uncovered in the literature. In this section we present our results in the nearly-critical regimes with critical Bond number ($B=1/3$).  Simulations were carried out  using different meshgrids and the results are independent of resolution.

\subsection*{Supercritical regime}
When only a single localised disturbance is considered, the supercritical flow has as its main feature the formation of an undular bore headed by a wave train with a train of waves propagating downstream (\cite{Milewski3}). 
\begin{figure}[!ht]
	\centerline{\includegraphics[scale=0.7]{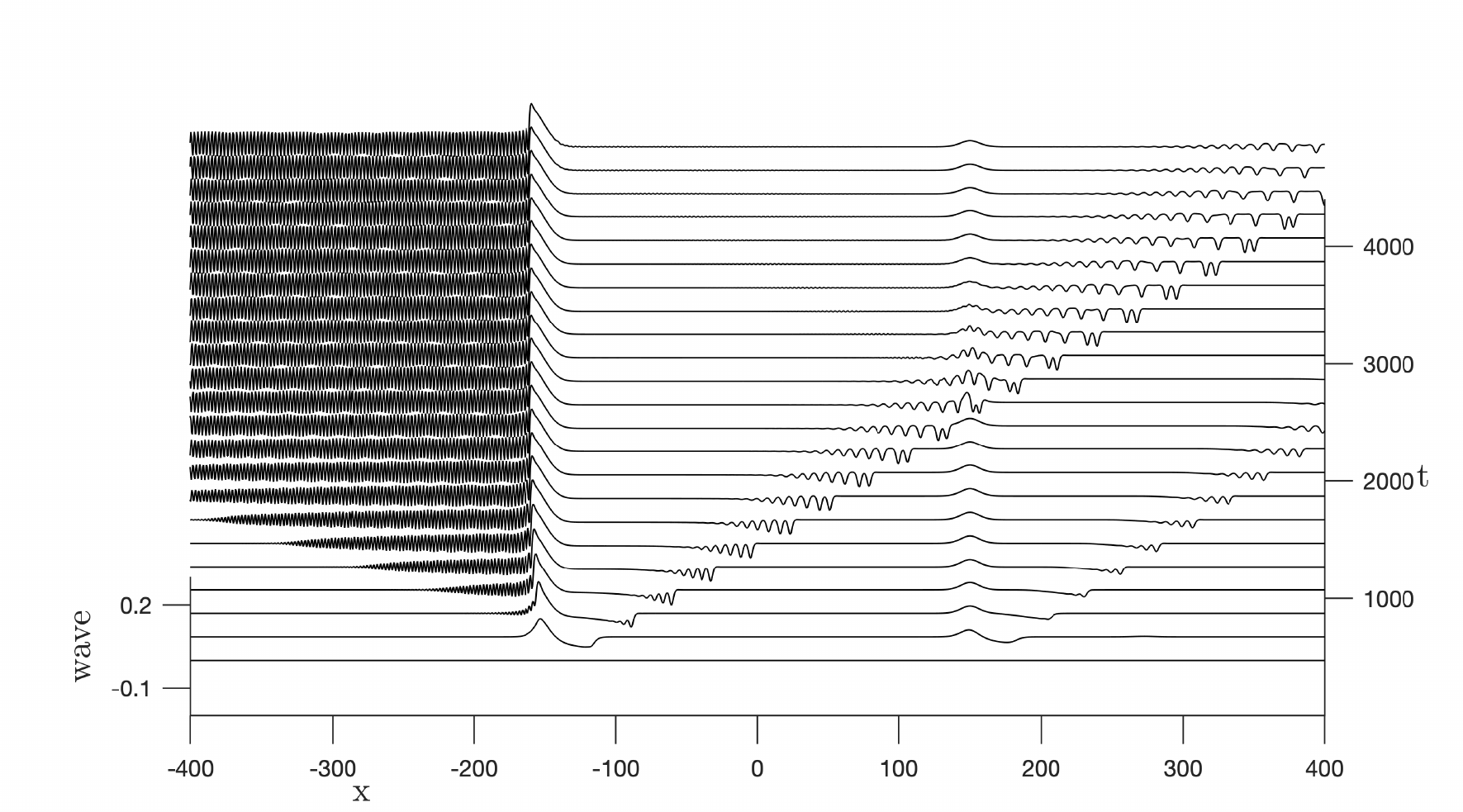}}
	\caption{Supercritical near-critical regime free-surface evolution with  $F=1.11$, $\epsilon_{1}=0.01$ and $\epsilon_{2}=0.005$. }
	\label{Fig1super}
\end{figure}

We start out our discussion considering the case in which the left disturbance has larger amplitude. For this purpose we fix $\epsilon_1 =0.01$ and $\epsilon_2 =0.005$. The dynamic resembles qualitatively the one reported by \cite{Capillary} (Figure 4.). We observe that waves are generated at each disturbances with depression solitary waves being emitted downstream from each one. While passing by the disturbance region, these waves radiate short waves with small amplitudes upstream. The radiation disturbs the elevation wave at the right disturbance and it is no longer a steady wave. This dynamic is show in details in Figure \ref{Fig1super}. 
\begin{figure}[!ht]
	\centering
	{\includegraphics[scale=0.7]{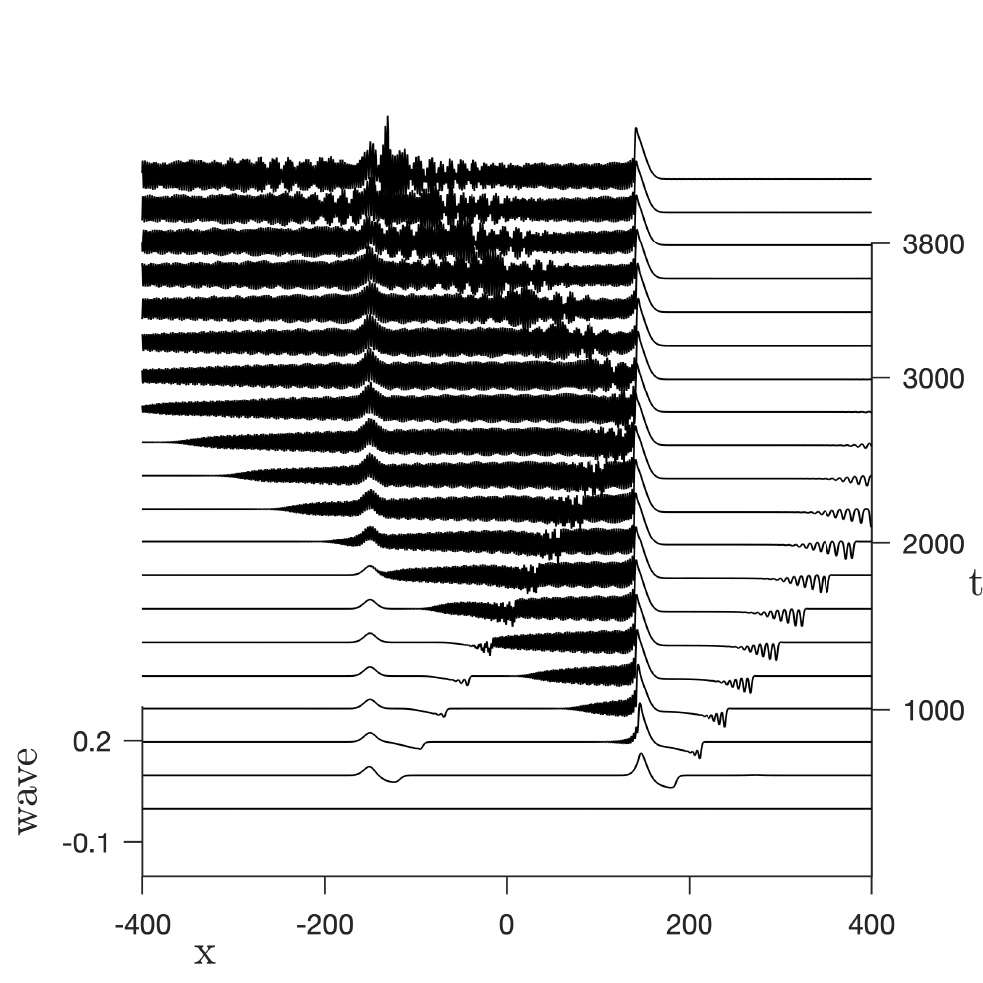}}
	{\includegraphics[scale=0.7]{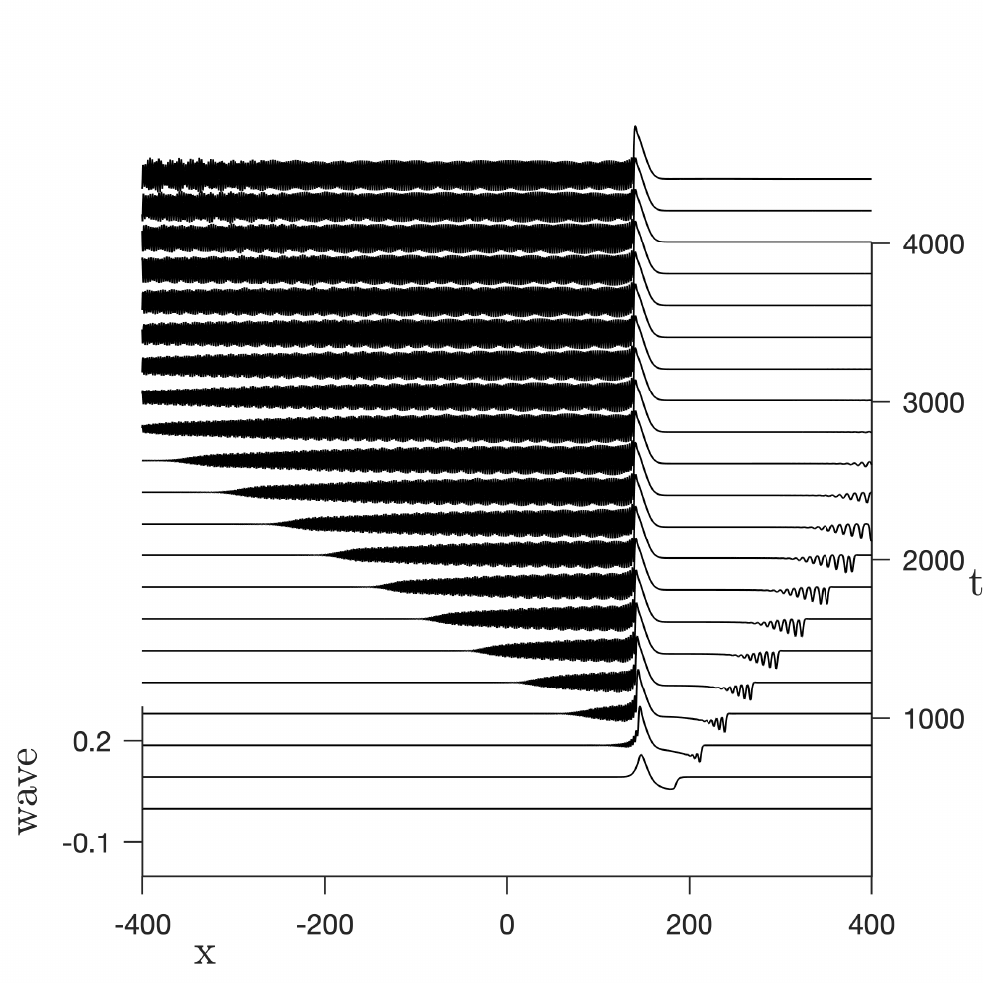}}
	\caption{Supercritical near-resonant regime free-surface evolution with  $F=1.11$. Left: $\epsilon_{1}=0.005$ and $\epsilon_{2}=0.01$. 
	Right: $\epsilon_{1}=0$ and $\epsilon_{2}=0.01$.}
	\label{Fig2super}
\end{figure}

Swapping the position of the disturbances,  at early times the dynamic is qualitatively similar to the one exhibited in Figure \ref{Fig1super}. However, for larger times the interaction in the neighbourhood where the weaker pressure is applied is somehow more nonlinear and we find numerical evidences that the wave may break during this interaction. This is shown in Figure \ref{Fig2super} (left). In order to verify that this possible wave breaking is due to the presence of the second disturbance, we turn off the weaker pressure and analyse the behaviour of the solution. Figure \ref{Fig2super} (right) describes this scenario. As it can be seen, when only one pressure is applied on the free surface the wave generated does not to break at any time. Although the forced fifth-order Korteweg-de Vries captures qualitatively the pattern of the generated waves (\cite{Capillary}) it fails to predict wave breaking, this phenomenon can only be  captured using the full nonlinear model.

In the same spirit, we maintain $\epsilon_1=0.01$ and  increase the amplitude of the right pressure, namely, $\epsilon_2=0.01$. As a consequence of that, wave breaking seems to occur at earlier times in a neighbourhood of where the left pressure is applied. This scenario is displayed in Figure \ref{e001e0005}.
\begin{figure}[!ht]
	\centerline{\includegraphics[scale=0.7]{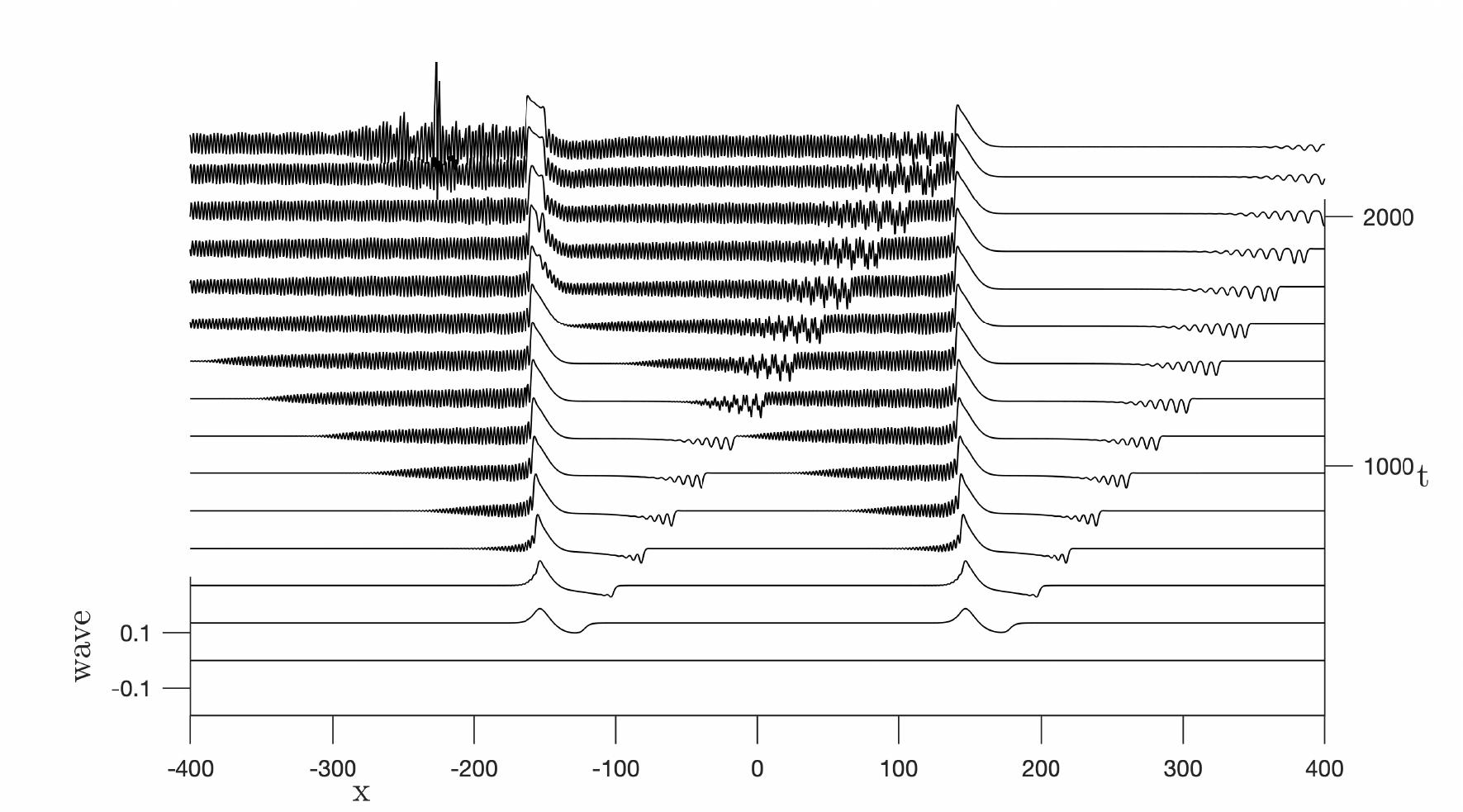}}
	\caption{Supercritical near-resonant regime free-surface evolution with  $F=1.11$, $\epsilon_{1}=0.01$ and $\epsilon_{2}=0.01$. }
	\label{e001e0005}
\end{figure}
\subsubsection*{Subcritical regime}
Considering a single localised disturbance in the fifth-order fKdV model, \cite{Milewski3} showed that the main feature of subcritical flows is the periodic generation of depression solitary waves which propagate downstream. Later, \cite{Capillary} found numerical evidences that when two localised obstacles are considered depression solitary waves may bounce back and forth  in the region between them remaing trapped for large times. Here, we confirm their prediction for the full nonlinear model. 

\begin{figure}[!ht]
	\centerline{\includegraphics[scale=0.7]{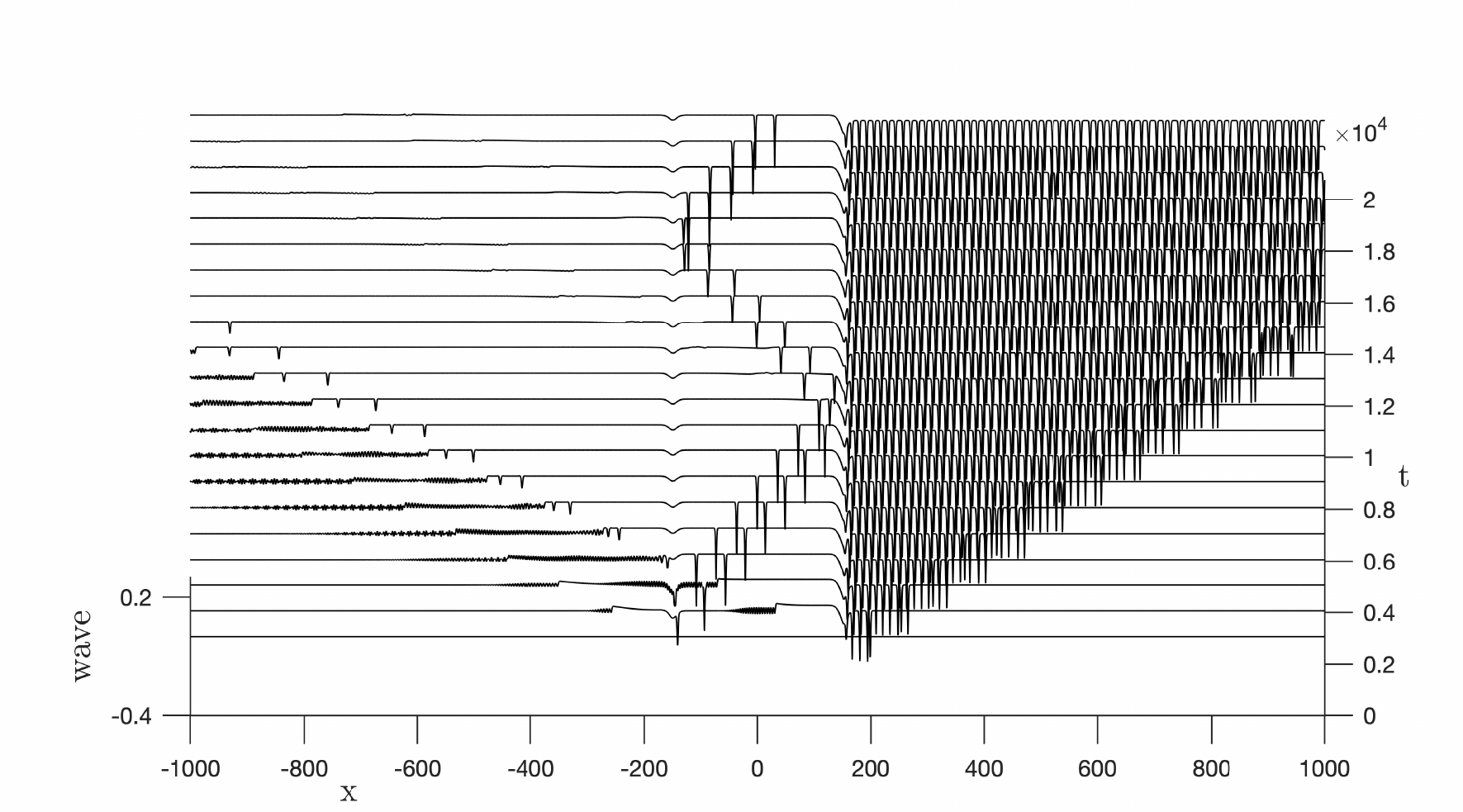}}
	\caption{Subcritical near-critical regime free-surface evolution with  $F=0.92$, $\epsilon_{1}=0.005$ and $\epsilon_{2}=0.01$. }
	\label{Fig1sub}
\end{figure}

For disturbances with different amplitudes, for instance $\epsilon_{1}=0.005$ and $\epsilon_{2}=0.01$, two trapped depression solitary waves can be seen in Figure \ref{Fig1sub} as well as their interactions. Although the wave interaction is almost linear,  it produces a nonlinear phenomenon, namely, trapping of waves.

\begin{figure}[!ht]
	\centering
	\includegraphics[scale=0.7]{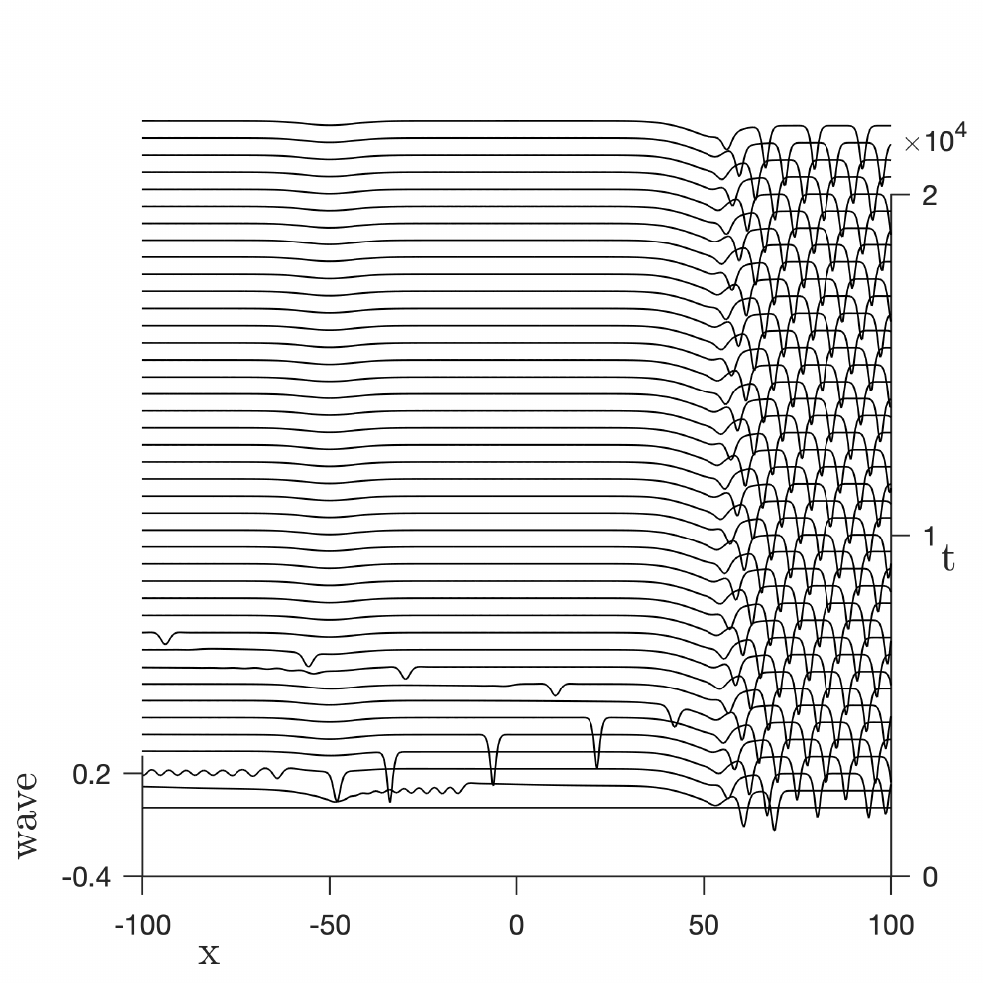}
	\includegraphics[scale=0.7]{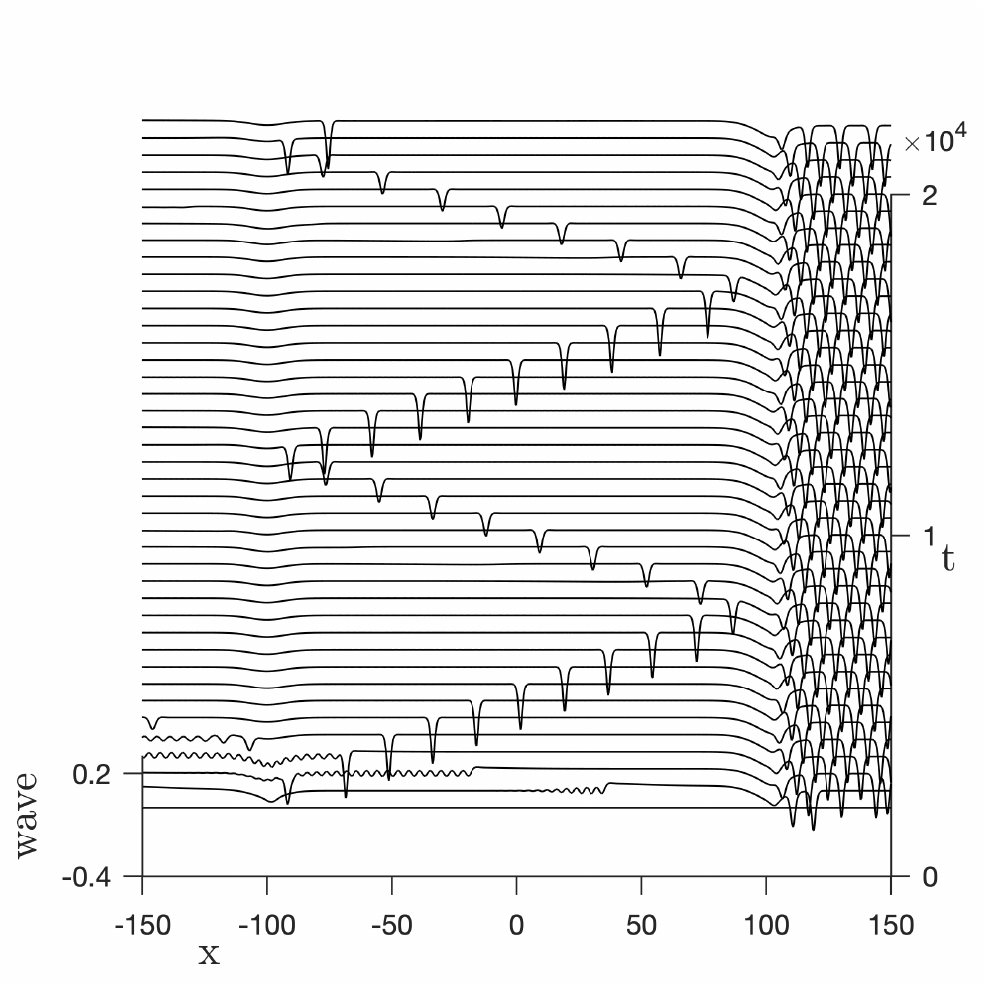}
	\includegraphics[scale=0.7]{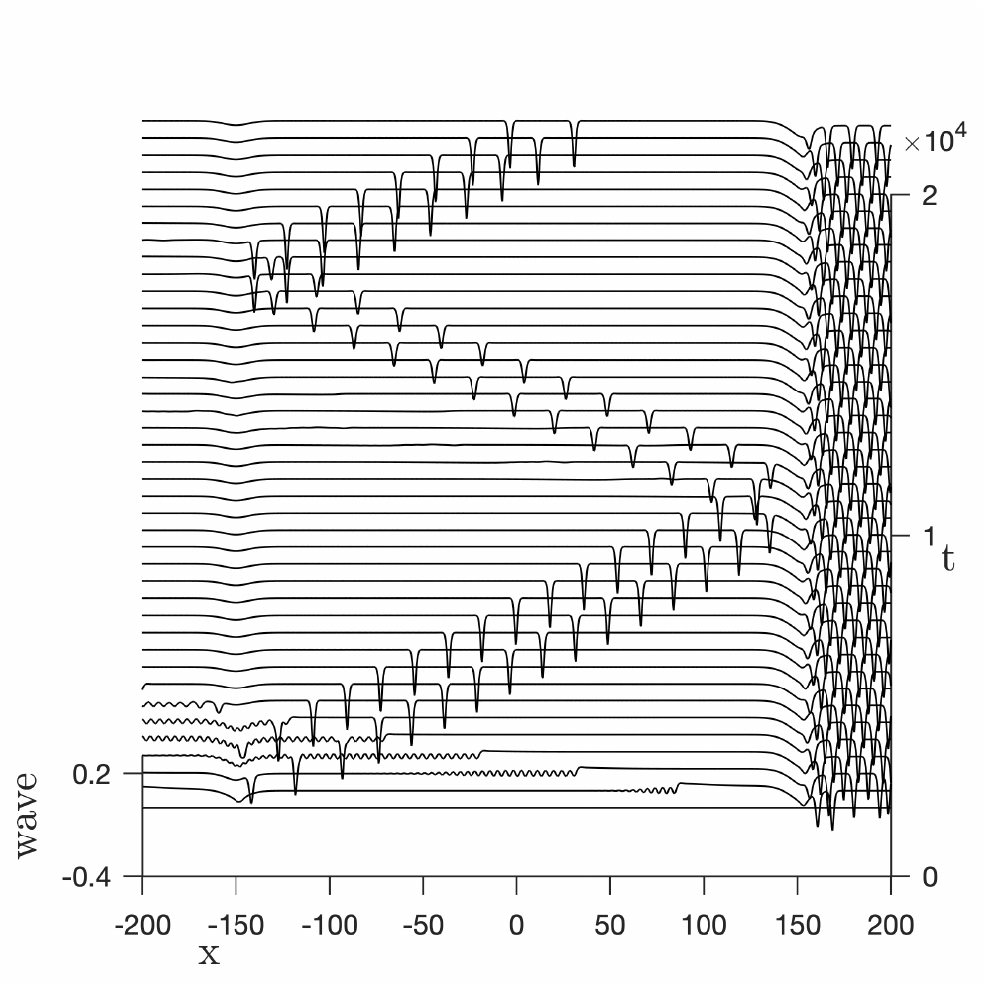}
	\includegraphics[scale=0.7]{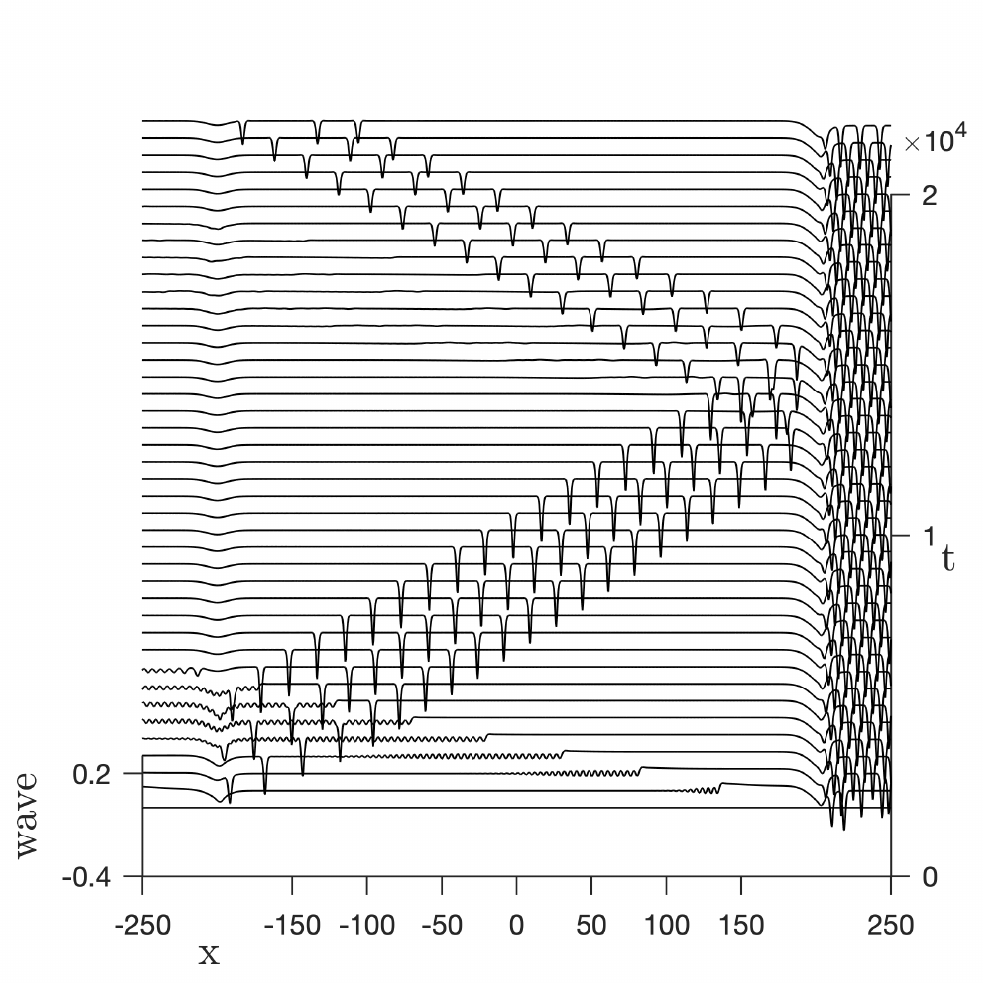}
	\caption{Trapped wave solutions between the disturbances with  $F=0.92$, $\epsilon_{1}=0.005$ and $\epsilon_{2}=0.01$. From top to bottom and left to right $d=100,200,300,400$. }
	\label{Fig2sub}
\end{figure}

Next we investigate the trapping mechanism by changing the distance of the disturbances. When they are closer the number of trapped waves between them decreases and when they are farther apart it increases. Figure \ref{Fig2sub} shows the dependence of the generation of trapped waves on the distance of the disturbances ($d=|x_a-x_b|$). When $d=0$ it is well known that there are not  trapped waves. However, when we increase the distance between the disturbances  trapped waves arise. 

\subsubsection*{Critical regime}
In the one-disturbance problem, the critical flow is mainly described by the formation of an undular bore where the pressure is applied  and a wave train propagating downstream (\cite{Milewski3}). 
\begin{figure}[!ht]
	\centerline{\includegraphics[scale=0.7]{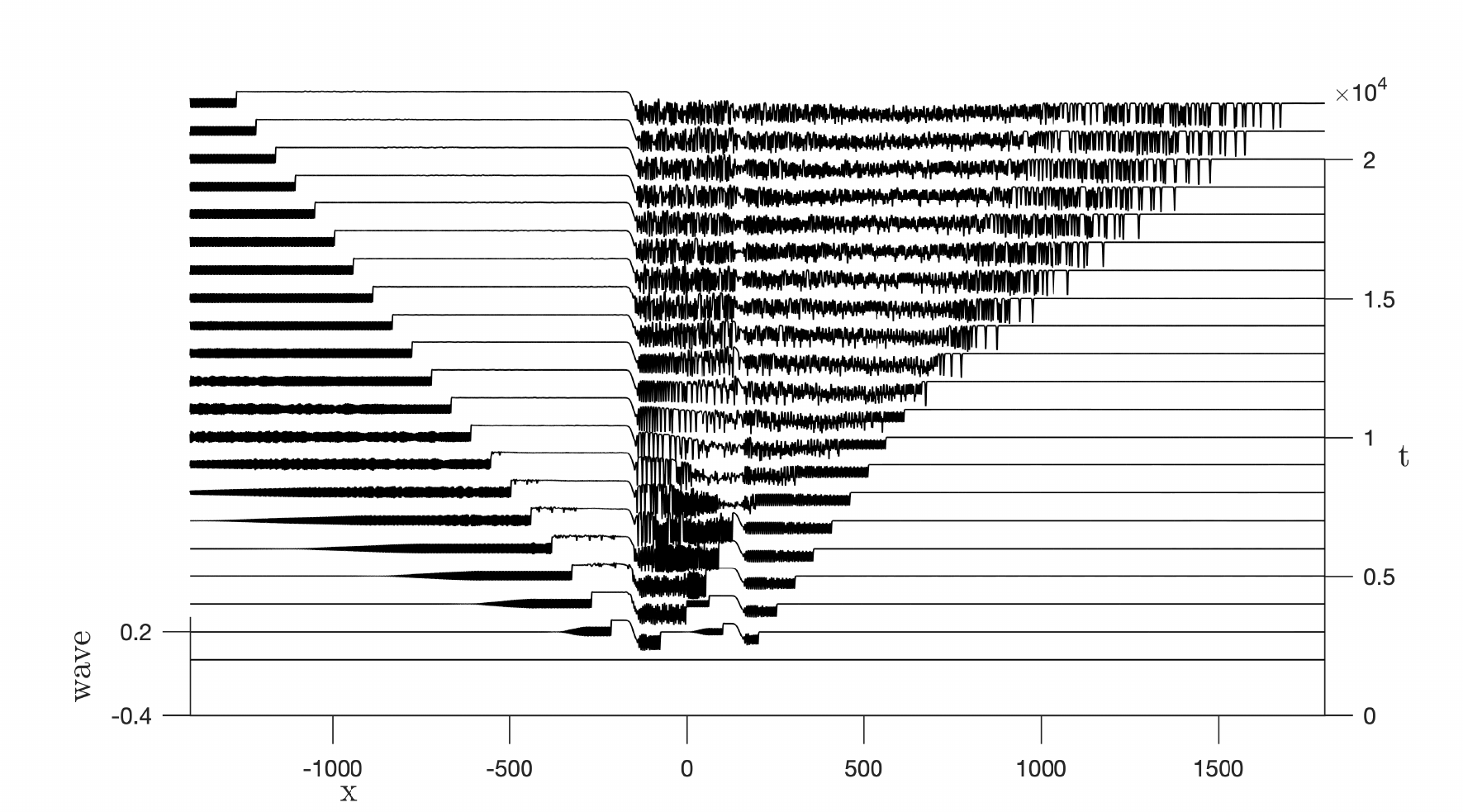}}
	\caption{Critical regime free-surface evolution with $\epsilon_{1}=0.01$ and $\epsilon_{2}=0.005$. }
	\label{Fig1crit}
\end{figure}

When two disturbances are considered our numerical simulations show that in certain regimes  depression solitary waves arise propagating downstream for large times. Figure \ref{Fig1crit} displays the free surface evolution in the critical regime for $\epsilon_1 =0.01$ and $\epsilon_2 =0.005$. At first we observe a dynamic similar to the one-disturbance problem, i.e., the formation of undular bores in the region where the pressures are applied with wave trains propagating downstream from each region. Once the waves start to interact, we notice that the flow is mainly controlled by the stronger disturbance, in the sense that the role of the weaker pressure cannot be point out in the motion. The novelty here is that well defined depression solitary waves emerge from the propagating downstream wave train for larger times. Under these circumstances we infer   that the critical regime captures features from the subcritical and supercritical regime simultaneously, i.e., the flow has depression solitary waves propagating downstream and   an undular bore at the disturbance headed by an upstream wave train.

\section{Conclusion}
In this article we have investigated capillary-gravity flows generated by the passage of two localised disturbances.  Through the use of the conformal mapping technique we find new features that were uncovered and not captured by fifth-order fKdV. Our numerical simulations show that in the supercritical regime the wave interaction is strongly nonlinear  which leads to an onset of wave breaking -- even when one of the disturbances has very small amplitude. In the subcritical regime, depression solitary waves remain trapped bouncing back and forth between the disturbances. In addition, we notice that more waves are trapped  as we increase the distance between the forcings. In the critical regime the presence of a second disturbance may generate depression solitary waves which propagate downstream. 
Besides, in this scenario we find that the critical regime carries features of the supercritical and subcritical flows simultaneously.

\section{Acknowledgements}

The authors are grateful to IMPA-National Institute of Pure and Applied Mathematics for the research support provided during the Summer Program of 2020. M.F. is grateful to Federal University of Paran{\' a} for the visit to the Department of Mathematics. R.R.-Jr is grateful to University of Bath for the extended visit to the Department of Mathematical Sciences.

\bibliographystyle{abbrv}

\end{document}